\documentclass[journal,11pt,onecolumn,final]{IEEEtran}
\usepackage{amssymb}
\usepackage{amsmath}
\usepackage{amsfonts}
\usepackage[doublespacing]{setspace}
\usepackage{geometry}
\usepackage{graphicx}
\usepackage{cite}

\begin{document}
\voffset=-5mm
\font\srm=cmr10 at 10pt
\vsize=9.4in

\def\svs{\vskip 3mm}\def\xvs{\vskip 2mm}
\def\shs{\hskip 8mm\relax}
\def\dhs{\hskip 4mm\relax}
\def\sn{\svs\noindent }\def\mn{\mvs\noin}\def\bn{\bvs\noin}
\def\mvs{\vskip 6mm} \def\mhs{\hskip 15mm\relax} 
\def\bvs{\vskip 10mm} \def\bhs{\hskip 20mm\relax} 
\def\noin{\noindent} 
\def\sqr#1#2{{\vcenter{\vbox{\hrule height.#2pt
\hbox{\vrule width.#2pt height#1pt \kern#1pt\vrule width.#2pt}
\hrule height.#2pt}}}}
\def\square{\mathchoice\sqr34\sqr34\sqr{2.1}3\sqr{1.5}3}
\def\cent{\centerline} 

\def\n{\noin}

\def\al{\alpha }\def\Al{\Alpha }\def\be{\beta }\def\ep{\epsilon }
\def\et{\eta }\def\sig{{\sigma }}\def\F{\Phi }\def\Up{\Upsilon }
\def\Del{{\Delta } }\def\del{{\delta} }\def\lam{{\lambda } }
\def\nab{{\nabla } }\def\Lam{{\Lambda }}\def\om{\omega }\def\gam{\gamma }
\def\Gam{\Gamma }
\def\th {\theta }

\def\bull{\vrule height 1.2ex width 1.0ex depth -.1ex } 
\def\den{\buildrel \triangle \over = } 
\def\hf{{1\over 2 }}\def\inta{\int_{\tau_k}^{\tau_{k+1}}}
\def\intu{\int_{\cal U} }\def\inth{\int_{t_0}^{t_0+h} }
\def\intth{\int_t^{t+h} } \def\nut{\nu (dt,du) }
\def\nutk{N(\del_{k+1},du)} \def\nutn{\intk\intu N(dt,du)}
\def\intk{\int_{t_k}^{t_{k+1}} }\def\da{{\downarrow} }
\def\inst{\int_s^t } 
\def\Be{{\cal B }}\def\A{{\cal A }}
\def\ve{\vert }\def\Ve{\Vert }
\def\L{{\cal L }}\def\D{{\cal D }}
\def\part{\partial} 
\def\intntt{\int_0^T } 
\def\intnt{\int_0^t }
\def\intns{\int_0^s } 
\def\intni{\int_0^{\infty} } 
\def\intnx{\int_0^x } 

\def\ov{\over}

\def\ah{{\hat a}}\def\sigh{{\hat \sigma}}\def\ch{{\hat c}}
\def\xh{{\hat x}}\def\xxh{{\bf{\xh}}}\def\aah{{\bf{\ah}}}
\def\cch{{\bf{\ch}}}\def\siggh{{\hat\sigg}}

\def\bl{\bigl}\def\br{\bigr}\def\Bl{\Bigl}\def\Br{\Bigr}
\def\bve{\big\ve }\def\Bve{\Bigr\ve }\def\bbve{\biggr\ve }
\def\BBve{\Biggr\ve }

\font\tenrfont=cmmib10
\newfam\mybm
\textfont\mybm=\tenrfont
\def\bsigma{{\fam\mybm \mathchar"711B}}
\def\sigg{\bsigma }
\def\sigb{\bar\sigg }

\def\bro{\bar\rho}
\def\X{\bar X}
\def\y{\bar y}
\def\x{\bar x}
\def\y{\bar y}
\def\p{\bar p}
\def\q{\bar q}
\def\U{\bar U}
\def\w{\bar w}
\def\d{\bar d}
\def\V{\bar V}

\def\B{\tilde B}
\def\A{{\cal A }}
\def\W{{\cal W }}
\def\C{{\cal C }}
\def\P{\tilde P}
\def\E{\tilde E}

\def\f{\phi }

\def\intf{\int_{-\infty}^{\infty}}

\def\nhs{\hskip -3mm}

\font\drm=cmr8 at 8pt
\font\drm=cmr8 at 8pt
\font\magitwo=cmti10 at 12pt

\font\itbf=cmbxti10

\def\bff{{\it\F}}
\def\bbf{{\bar{\hskip -1mm {\it\F}}}}
\def\ibF{{\it \bar \F}}
\def\vsn{\vskip -3mm} 
\def\bs {\boldsymbol}

\title{Exact Amplitude Distributions of Sums of Stochastic Sinusoidals and
their Application in Bit Error Rate Analysis}

\author{Y. Maghsoodi \\
Scinance Analytics, www.ScinanceAnalytics.com, \\
ym@ScinanceAnalytics.com\linebreak 
\and A. Al-Dweik, \textit{%
Senior Member, IEEE} \\
Khalifa University, P.O.Box 127788, Abu Dhabi, UAE\\
dweik@fulbrightweb.org\\
This version 2019\\
Copyright Y. Maghsoodi and A. Al-Dweik 2002}
\date{}
\maketitle

\vskip -15mm
\begin{abstract}
 
We consider a model applicable in many communication systems where the sum of 
$n$ stochastic sinusoidal signals of the same frequency, but with random amplitudes as well as 
phase angles is present. The exact probability distribution of the resulting 
signal's amplitude is of particular interest in many important applications, 
however the general problem has remained open. We derive new general formulae 
for the resultant amplitude's distribution, in terms of any given general joint 
density of the random variables involved. New exact probability densities of cases 
of interest follow from the general formulae  which are applied to the problem of the
exact evaluation of the bit error rate performance in narrowband multipath
fading channels that consist of a small number of jointly dependent
resolvable multipath components. The numerical results are compared 
with those of the currently popular method of Gaussian approximation.
These are consistent with simulations, and show significant increase in accuracy 
particularly for small number of components.\end{abstract}


\evensidemargin 0in\oddsidemargin\evensidemargin

\section{Introduction}

In many communication systems the received signal can be expressed as \cite{Proakis-1}-\cite{Kah-1} 
\begin{equation}
r(t)=\sum_{i=1}^{n}A_{i}\cos(\omega t+\phi_{i})+Z(t)\ .  \label{1.1}
\end{equation}
The first term in (\ref{1.1}) represents the desired signal component with
amplitude $A_{1}$, frequency $\omega$, and phase $\phi_{1}$, mixed with
interference that is produced either unintentionally by a multiple access
process, reflections of the desired signal, or intentionally by a certain
jammer. The second term, $Z(t)$, is an additive white Gaussian
noise (AWGN) with two-sided power spectral density of $N_{0}/2$.

In general the amplitudes $(A_{1},A_{2},...,A_{n})$, and the phases $(\phi_{1},\phi_{2},...,\phi_{n})$%
, may all be random variables, with a given joint probability distribution
function, which is the framework treated in this paper. In specific
applications however, the statistical properties of these variables, and the
value of $n$, depend on the particular channel model and the application considered. 
For example, in the case of narrowband
local area UHF and microwave propagation, the amplitudes are considered to
be constants and the phases are independent and uniformly distributed over $%
[-\pi ,\pi ]$ \cite{Proakis-1}-\cite{Rappaport-book}. The number of
multipath components $n$ is determined based on the terrain of the
surroundings at the receiver side. Similar assumptions are usually made for
synchronous frequency hopping spread spectrum systems (FH-SSS) with multiple
access interference (MAI) or multitone jamming (MTJ) with AWGN \cite{Kah-1}-%
\cite{Arafat-1}. The value of $n$ in multiple access systems corresponds to
the number of simultaneous users and it is modelled as a random variable
with binomial distribution.

The case where the amplitudes are random has also a wide range of
applications. For example, Beaulieu et al \cite{Beaulieu-1} have modelled the
amplitudes as a Nakagami-$m$ random variables to analyze the performance of
wireless networks with cochannel interference. The
Rayeligh-distributed amplitudes model was adapted for FH-SSS with orthogonal
frequency division multiplexing (OFDM) in \cite{Arafat-1}, \cite{Kim 1}.

In evaluation of the performance of communication systems with received
signals such as $r(t)$ in (\ref{1.1}), the knowledge of the pdf
(probability density function) of the resultant amplitude or envelope is of
crucial importance. Rice \cite{Rice-1} stated that the cdf (cumulative
distribution function) of the envelope of some special cases of (\ref{1.1})
may be represented by a Fourier-Bessel Transform. Kluyver \cite{Kluyver}
considered the case of $n$ randomly phased sine waves in the absence of
noise and gave a corresponding equation for the special case in \cite%
{Rice-1}. Simon \cite{Simon-1} considered a special recursive case of (\ref%
{1.1}) without noise, where all the amplitudes are fixed, and each new phase
angle is assumed to be equal to the resultant of all the previous ones, plus an
independent uniformly distributed angle. Thus, he deduced a recursive
formula for the pdf of the squared envelope in this special case (see also Section III below). 
Helstrom \cite{Helstrom-1} derived the amplitude pdf of the sum of two sinusoidals of constant
amplitude, affected by a Gaussian noise, and later developed a numerical
scheme to approximate the cdf of the envelope when $n>10$, in the presence
of a narrow band Gaussian noise \cite{Helstrom-2}. 
Beckman \cite{Beckmann-2}, considered and cited several other special cases 
of the problem posed in model (\ref{1.1}), including constant amplitudes with a non-uniform 
pdf for the $\phi_i$s, and $X$ and $Y$ having a non-Gaussian joint pdf, where
$X\den \sum_{i=1}^{n}A_{i}\cos(\phi_{i})$, and $Y\den \sum_{i=1}^{n}A_{i}\sin(\phi_{i})$.
The latter special case was also considered by Zabin and Wright \cite{Zabin}. 
Abdi et al. \cite{Abdi} assumed independent and identically uniformly distributed phases, 
with arbitrarily dependent amplitudes 
and derived a multiple integral formula for the envelope pdf, for which they also gave 
an infinite Laguerre expansion. They applied their results
to study the statistical behaviour of the scattering cross
section when the number of scatters is small and deterministic and all
amplitudes are equal. As also reported in \cite{Beckmann-2} and \cite{Abdi}, in most practical cases 
the uniform pdf assumption for the phases is usually satisfied.  
Other authors too, have made special assumptions of various forms, 
about the distributions and independence of the phases and the amplitudes. A brief survey of these works 
can also be found in Abdi et al. \cite{Abdi}.  Maghsoodi \cite{Maghsoodi-1} derived exact formulae 
for the envelope pdf for the general case, where the amplitudes and phases were not assumed to be 
independent,  nor were they assumed to have any particular probability distribution.

In this paper, we expand the work in \cite{Maghsoodi-1} and discard the restrictive assumptions other previous works, 
and present general formulae for the envelope pdf in the most general case. We then apply the results 
to bit error rate analysis. The extension of our formulae to include random number of signals is immediate under 
independence, and still follows naturally under no independence, though slightly less immediately.

We first consider model (\ref{1.1}) in the absence of noise, 
and while allowing all the amplitudes as well as phases to be random, 
having a given general joint pdf, we derive two exact general formulae for the 
probability distribution of the envelope. These are proved in 
Theorems 1 and 2 \cite{Maghsoodi-1}. The presented formulae will be solely
in terms of integrals of any given general joint probability density of all
the amplitudes and phases in the sum. Examples of the implementations of the
general formulae, yield interesting new exact densities of some important
cases of interest. These are applied to evaluate the exact error probability of binary phase
shift keying (BPSK) systems over narrowband fading channels that consist of
a small number of multipath components. It turns out that significant gain in 
accuracy is made, particularly over the currently popular method of Gaussian approximation, where
the application of the Central Limit Theorem is much less accurate for smaller number of components. Examples for using the proposed approach for evaluating the error probability can be found in \cite{Arafat-1, Arafat-2, Arafat-3}.

The paper is organized as follows. In section II Sinusoidals Addition 
Theorem (SAT), cites the formulae for the resultant amplitude and phase 
angle of sums of sinusoidals. Allowing for random amplitudes and phase angles, Theorem 1 
presents EDDHAPT (Envelope Distribution from Density of 
Half Angle Phase Tangents) formula, for the exact cdf of the envelope, in 
terms of the joint pdf of the amplitudes and the tangents of the half phase angles. Two 
examples of the implementation of this formula will follow. In section III,  
Envelope Separation Theorem (EST) gives a recursive formula for the envelopes, followed by Theorem 2 
which presents the EGED (Exact General Envelope Density) formula \cite{Maghsoodi-1}, which allows the calculation of the pdf of the 
envelope, in terms of the given joint pdf of the amplitudes and the actual phase 
angles themselves. In section IV we apply
the results of the preceding sections to evaluate the exact error
probability of BPSK systems in narrowband multipath fading channels.
Finally, the conclusions are presented in Section V and the mathematical
proofs are given in the appendix.

\section{New formulae for the envelope and its cdf via half-phase tangents}

In this section we assume that the joint pdf of all the amplitudes and the 
tangents of all the half phase angles of model (\ref{1.1}) is given. The use of this joint pdf is 
more convenient here due to the range $%
(-\infty,\infty)$ of the tangent. There is no loss of generality in this assumption, 
since this joint pdf can always be obtained from that of the amplitudes and 
the phase angles themselves by suitable transformations. In what follows 
we shall denote by $|\mathbf{A}|$ the Euclidean norm of the vector $(A_{1},...,A_{n})$.

\vskip 4mm 

The SAT states that, given the identity  \begin{equation}
\sum_{i=1}^{n}A_{i}\cos (\omega t+\phi _{i})=B_{n}\cos (\omega t+\theta _{n})%
\text{\shs }t\geq 0\ ,  \label{E-2.1}
\end{equation}

\n where $A_{i}\in (-\infty ,\infty )${\ and }$\phi _{i}\in
\lbrack -\pi ,\pi ]${, there exists a unique solution of \textrm{(\ref {E-2.1})} for }$\{B_{n},\theta _{n}\}${\ which is independent of }$%
\omega ${\ and }$t${\ and is given by  }%
\begin{equation}B_{n}^{2}=|\mathbf{A}|^{2}+2\nhs\sum_{n\geq j>k\geq 1}\nhs A_{j}A_{k}\cos
\left( \phi _{j}-\phi _{k}\right)  \label{2.2}
\end{equation}%
\begin{equation}
\theta _{n}=\tan ^{-1}{\frac{\sum_{i=1}^{n}A_{i}\sin \left( \phi _{i}\right) 
}{\sum_{i=1}^{n}A_{i}\cos \left( \phi _{i}\right) }}\ ,  								\label{E-2.3}
\end{equation}%
{where }$\theta _{n}${\ can be uniquely chosen such that }$%
B_{n}${\ is always positive.}  The proof is by writing (\ref{E-2.1}) with $\sin$ as well,  and expanding and solving 
the resulting double identities for $(B_{n},\theta _{n})$. 

It follows from formula (\ref{2.2}) ( see also (\ref{3.1}) below and \cite%
{Maghsoodi-1}) that if $\phi_{n}$ can take all possible values in $%
[-\pi,\pi] $, then 
\begin{equation}
\left\vert B_{n-1}-|A_{n}|\right\vert \leq B_{n}\leq B_{n-1}+|A_{n}|     						\label{E-2.5}.
\end{equation}
Hence, we can have a recursion for the minimum and maximum possible values
of $B_{n}$, in terms of those of $B_{n-1}$, which we denote by $m_{n}$ and $M_{n}$ 
respectively. In each specific application the values of $m_{n}$
and $M_{n}$ would strictly depend on the specific range of the values of $A_{i}$
and $\phi_{i}$ for $i=1,2,...,n$, however the above inequalities would
determine $m_{i}$ and $M_{i}$ for $i=1,2,..,n$ recursively in each case. For
example when $A_{1}$ and $A_{2}$ are constants, $B_{1}=|A_{1}|$, $%
M_{2}=|A_{1}|+|A_{2}|$ and $m_{2}=\left\vert |A_{1}|-|A_{2}|\right\vert $,
and so on.

In what follows unless otherwise specified, upper case
letters such as $A_{1}$ and $T_{1}$ will denote random variables and lower
case letters such as $a_{1}$ and $t_{1}$ will represent their corresponding
possible values. In addition, bold-face letters such as $\mathbf{A}$ will
denote vectors with elements in regular fonts such as $\mathbf{A}%
=(A_{1},...,A_{n})$, and $d\mathbf{a}$ will denote $da_{1}da_{2}...da_{n}$
and so on. Further, $\bar{\mathbf{T}}$ and $\bar{\mathbf{t}}$ will denote
the truncated vectors $(T_{1},...,T_{n-1})$ and $(t_{1},...,t_{n-1})$
respectively.

\vskip 3mm\textbf{Theorem 1} ( The EDDHAPT Formula) \textit{Consider the sum of }$n$\textit{\ stochastic sinusoidals on the
LHS of \textrm{(\ref{E-2.1})}, with }$A_{i}\in(-\infty,\infty )$, \textit{\
and }$\phi_{i}\in\lbrack-\pi,\pi]$, \textit{\ random variables with a given
general joint pdf }$f_{_{\mathbf{A},\mathbf{T}}}(\mathbf{a},\mathbf{t}%
)\triangleq f_{A_{1}...A_{n},T_{1}...T_{n}}(a_{1},...,a_{n},t_{1},...,t_{n}),$%
\textit{\ where }$T_{i}\triangleq\tan(\phi_{i}/2), i=1,2,\cdots,n$\textit{. Then the cdf of
the resultant amplitude }$B_{n}$\textit{\ is given by}%

\vskip -6mm

\begin{equation}
P(B_n \leq b_n) =\int_{-\infty}^\infty d{\bf a} \int_{-\infty}^{\infty} d\bar {\bf t} {\bf 1}_{\{ a<0\}} f_{{\bf A},{\bf T}} ( {\bf a, \bar t}) +  
\int_{-\infty}^{\infty}  d{\bf a} \int_{-\infty}^{\infty}  d\bar {\bf t} {\bf 1}_{\{\Delta_Q >0 \}}  \int_{\bar a_n}^{\bar b_n}dt_n f_{{\bf A},{\bf T}} ( {\bf a, t})   			\label{2.6}
\end{equation}

\n {\it where}\ \ $\Delta_{Q}\triangleq4a_{n}^{2}\bar{S}_{n}^{2}-a(I_{n}+2a_{n}\bar{C}%
_{n})$, \ \ \ \ $\bar{S}_{n}\triangleq\sum_{i=1}^{n-1}\frac{2a_{i}t_{i}}{%
1+t_{i}^{2}}$, \ \ \ \ $\bar{C}_{n}\triangleq\sum_{i=1}^{n-1}\frac {%
a_{i}(1-t_{i}^{2}{)}}{1+t_{i}^{2}}$, \ \ \ \ $a\triangleq I_{n}-2a_{n}\bar {C%
}_{n}$

\svs
$I_{n}=\sum_{i=1}^{n}a_{i}^{2}+\bar{k}_{n}-b_{n}^{2}, \ \ \bar{k}_{n}= 2 \sum_{(n-1)\geq j>k\geq1}\nhs\ \ {{ a_{j}a_{k}( (1+t_{j}t_{k})^{2}}-(t_j-t_k)^2)\ov \left ( 1+t_{j}^{2}\right) \left( 1+t_{k}^{2}\right) },
\ \ \ \bar{a}_{n},\bar{b}_{n}\triangleq\lbrack{-2a_{n}\bar{S}_{n}\mp\sqrt{{\Delta}_{Q}}}]/{a},$ 

\vskip 2mm
\noin\textit{ and ${\bf 1}_{\{\cdot\}}$ denotes the indicator function of the set $\{\cdot\}$}.

\mn\textit
{Example 1}\ \ \ As an example of the implementation of the EDDHAPT formula (\ref {2.6}) 
assume that $n=2$, $A_{1}=A_{2}=A$, a constant, and $\phi_{1}$ and $%
\phi_{2}$ are independently uniformly distributed in $(-\pi,\pi)$. It can
then be easily deduced that the cdf $F_{T_{i}}(t_{i})$ and the pdf $%
f_{T_{i}}(t_{i})$ of $T_{i}$ are respectively given by \cite{Maghsoodi-1} 

\svs\shs $F_{T_{i}}(t_{i})={\frac{1}{\pi}}[\tan^{-1}(t_{i})+{\frac{\pi}{2}}]$\dhs and \dhs $%
f_{T_{i}}(t_{i})=\frac{1}{\pi(1+t_{i}^2)}$\dhs  $\ \ t_{i}\in(-\infty
,\infty),\ \ i=1,2$.

\sn Application of the EDDHAPT formula (\ref {2.6}) immediately gives $\Delta_Q=b_2^2(4A^2-b_2^2)$ 
which is always positive since $b_{2}\in (0,2\vert A\vert)$ and $a=[(4A^{2}-b_{2}^{2})t_{1}^{2}-b_{2}^{2}]/(1+t_{1}^{2})$. Hence 

\vskip -5mm
\begin {align} P(B_2 \leq b_2)
&=\int_{-\infty}^\infty {\bf 1}_{\{a<0\}}\ dt_1f_{T_1}(t_1) + \int_{-\infty}^\infty dt_1 \int_{\bar a_2}^{\bar b_2} dt_2 f_{T_1}(t_1)f_{T_2}(t_2)\cr
&={1\ov \pi} \int_{-\infty}^\infty {\bf 1}_{\{\ve t_1\ve>\alpha\}} {dt_1\ov 1+t_1^2} + {1\ov \pi^2} \int_{-\infty}^\infty dt_1 {{\cal T}(\bar a_2,\bar b_2)\ov 1+t_1^2}	\label{2.7}
\end{align}						
\vskip -3mm\n where $\bar{a}_{2},\bar{b}_{2}=\frac{-4A^{2}t_{1}/(1+t_{1}^{2})\mp\sqrt {%
\Delta_{Q}}}{a}$, ${\cal T}(\bar a_2,\bar b_2)\den \tan^{-1}(\bar b_2)-\tan^{-1}(\bar a_2)$, 
and $\alpha=\frac{b_{2}}{\sqrt{4A^{2}-b_{2}^{2}}}$ is the positive root of $a=0$.
Integrating the second integrand by parts, whilst noting the singular discontinuities of $\bar{a}_{2}$ and $\bar{b}_{2}$%
, at $t_{1}=\pm\alpha$, and the fact that 

\vskip -3mm
$$\bar a_2'(1+\bar a_2^2)^{-1}=\bar b_2'(1+\bar b_2^2)^{-1}=(1+t_1^2)^{-1}\ ,$$

\vskip -4mm
\n we obtain 

\vskip -6mm
\begin{equation}
P(B_{2}\leq b_{2})={\frac{2}{\pi}}\tan^{-1}{\frac{b_{2}}{\sqrt{%
4A^{2}-b_{2}^{2}}}}\ ,  															\label{2.8}
\end{equation}
which after differentiation gives the pdf of the amplitude%
\begin{equation}
f_{B_{2}}(b_{2})=\left\{ 
\begin{array}{cc}
\frac{{2}}{{\pi\sqrt{4A^{2}-b_{2}^{2}}}} & \text{ \ \ \ }0\ \leq{b}_{2}{\
\leq2|A|} \\ 
0\text{ \ \ \ \ \ \ \ } & \text{otherwise}\ .%
\end{array}
\right. 												 					\label{2.9}
\end{equation}
(See Figure 1 for a graph of this pdf).

\mn\textit
{Example 2}\ \ \ Assume that in example 1 above,  $\phi_1$ and $\phi_2$ are dependently distributed with joint pdf

\vsn\begin{equation}
f(\phi_1,\phi_2)={1\ov \pi^3}(\phi_1+\phi_2)\shs         0\le \phi_i\le \pi .						\label{2.15}
\end{equation}

\n It can easily be verified that the joint pdf of the corresponding $T_i$ random variables is \cite{Maghsoodi-1} 
\begin{equation}
f_{T_1,T_2}(t_1,t_2)={8(\tan^{-1} t_1+ \tan^{-1} t_2)\ov \pi^3 (1+t_1^2)(1+t_2^2)}             \shs         0\le t_i< \infty\ ..\label{2.16}
\end{equation}

\n Then, in this case formula (\ref{2.6}) above reduces to 

\begin {equation} 
P(B_2 \leq b_2)=\int_{-\infty}^\infty {\bf 1}_{\{\ve t_1\ve>\al\}}\ dt_1 \int_{-\infty}^\infty  dt_2 f_{T_1,T_2}(t_1,t_2) 
+ \int_{-\infty}^\infty dt _1 \int_{\bar a_2}^{\bar b_2}  dt_2 f_{T_1,T_2}(t_1,t_2).					\label{2.17}
\end{equation}						

\n where the parameter values are as in example 1, except the joint pdf which is given in (\ref {2.16}).  Calculating the integrals in (\ref {2.17}), by methods very similar to example 
1 above we obtain 
\begin{equation}
F_{B}(b)={4\ov \pi^2} (\tan^{-1}\alpha)^2  \shs  0\le b\le 2\vert A\vert\ .\label{2.18} 
\end{equation}	
\n Hence, the amplitude pdf for this example is  
\begin{equation}
f_{B}(b)={8\ov \pi^2} {\tan^{-1}\alpha\ov\sqrt{(4A^2-b^2)}}  \shs  0\le b\le 2\vert A\vert\ . 							\label{2.19}
\end{equation}
\n From the pdf (\ref{2.19}) it is evident that it has a singularity at $b_2=2A$ which can also be seen in Fig. 1, where the pdf is illustrated for $A=1$.

Example 2 also illustrates that formula (\ref{2.6}) can find the envelope pdf in the more general, and practically more realistic situations where the phase angles may have any arbitrary joint law, and may not be limited to be independent or be uniform. This example can have applications where the phase angles are correlated, and their probability of occurrence would linearly increase with the angle, e.g. in signal propagation over correlated multipath fading channels, particularly the 2-ray models used for mobile radio channels.

\section{Envelope separation and its pdf in terms of the joint pdf}

In this section we represent the pdf of the envelope, directly in terms of any given joint pdf of the amplitudes and the phase
angles in the stochastic sinusoidals sum. First the Envelope Separation 
Theorem (EST) is presented, which gives a recursive formula for the $n^{th}$ envelope in
terms of the $(n-1)^{st}$ envelope and resultant phase angle. This can also be viewed 
as a type of second cosine theorem. Theorem 2 then derives a new Exact General 
Envelope Distribution (EGED) formula for the envelope pdf. The advantage of this formula over (\ref{2.6}) is that the
regions of integration are explicitly determined, however the disadvantage
in applications is the complications of dealing with bounded regions of
integrations compared to $(-\infty,\infty)$ of $t_{i}$ in (\ref{2.6}).

\sn The EST expresses the $n^{th}$ envelope $B_{n}^{2}$  as
\begin{equation}
B_{n}^{2}=B_{n-1}^{2}+A_{n}^{2}+2A_{n}B_{n-1}\cos(\phi_{n}-\theta_{n-1}).
\label{3.1}
\end{equation}

\n The proof is by separating all the terms involving $A_n$ in formula  (\ref{2.2}), and writing the remaining 
terms as $B_{n-1}^2$ \cite{Maghsoodi-1}.

It can immediately be seen from (\ref{3.1}) that, if we assume $\phi_{i}$s 
to have a special form such that, $\{\xi_{i}\triangleq\phi_{i}-%
\theta_{i-1},\ i=2...n,\ \ \theta_{1}=\phi_{1}\}$ is an i.i.d. (independent and identically  distributed) 
sequence of random variables, then clearly the sequence $%
\{B_{i}^{2},\ i=1,2,...\}$ becomes Markovian, and a simple recursive formula
can be written for its distribution. This assumption was made by Simon \cite%
{Simon-1}, where the $\xi_{i}$s were further assumed to be independently
uniformly distributed. Though Simon's work is interesting, and his formula
can be derived as special cases of
the results of this paper, however the assumptions are very restrictive on
the general model considered here as well as on the scope of practical
applications, in particular these assumptions imply that, each received sinusoidal adapts
its phase to the resultant phase of all the previous sinusoidals in a
special way, which also leads to a particular joint law for the $\phi_{i}$s. Since 
by our formula (\ref{E-2.3}), Simon's assumption is equivalent to saying that, $\{\phi_i,\ i=1\cdots,n\}$ are such that, 
they satisfy 
$$\xi_{i}\den \phi_i-\tan ^{-1}{\frac{\sum_{j=1}^{i-1}A_{j}\sin \left( \phi _{j}\right) 
}{\sum_{j=1}^{i-1}A_{j}\cos \left( \phi _{j}\right) }}\sim U(-\pi,\pi)\mhs i=2,\cdots,n$$
\n and the $\xi_{i}$\ s are i.i.d., which are complicated special modelling assumptions about the $\{\phi_i,\ i=1\cdots,n\}$.  
Another immediate consequence of Simon's assumption is that, the phases can never be independent, since, 
for example under this assumption we would have 
\begin {align} P(\phi_2\le y/\phi_1=x)=
&P(\phi_2-\phi_1\le y-x/\phi_1=x)\cr=
&{{y-x+\pi\ov 2\pi}}\cr\neq 
&P(\phi_2\le y)\ .								\nonumber
\end {align} 
But, alternatively, in the absence of any additional (physical) information to the contrary, it would be more natural and physically meaningful, to assume that each signal's phase is independent of the others, and is uniformly distributed. Our general formulae allow us to model and solve these cases as well, as special cases. Moreover, at the expense of little extra mathematical complexity, we can still derive recursive formulae in the general independent cases without Simon's assumptions.

\svs\textbf{Theorem 2} (Exact General Envelope Density (EGED) via the joint pdf )
\textit{\ \ Under the assumptions of Theorem 1, if the amplitudes and phases random 
variables $\mathbf{A}$ and $\mathbf{\Phi}$, have a given general joint pdf }
$f_{_{\mathbf{A},\mathbf{\Phi}}}(\mathbf{a},\mathit{\Phi})\triangleq
f_{A_{1}...A_{n},\Phi_{1}...\Phi_{n}}(a_{1},...,a_{n},\phi_{1},...,\phi_{n})$%
\textit{, then the pdf of the envelope }$B_{n}$\textit{\ is given by }
\begin{equation}
f_{B_{n}}(b_{n})=2b_{n}E\left\{ \left\vert \sqrt{%
4A_{n}^{2}B_{n-1}^{2}-(b_{n}^{2}-B_{n-1}^{2}-A_{n}^{2})^{2}}\right\vert
^{-1}1_{\{\left\vert \Psi_{n}\right\vert \le 1,\ \Phi_{n}\in U_{n}\}}\right\} 
\text{ \ \ }m_{n}\le b_{n}\le M_{n}\ ,  						\label{3.3}
\end{equation}
\n\textit{\ where the expectation is with respect to the given joint law }$%
f_{_{\mathbf{A},\mathbf{\Phi}}}$\textit{, and }$\mathbf{1}_{\{.\}}$\textit{\
denotes the indicator of the set }$\{.\}$\textit{\ and}%
\begin{equation}
\mathbf{\Psi}_{n}\triangleq\left( \frac{b_{n}^{2}-B_{n-1}^{2}-A_{n}^{2}}{%
2A_{n}B_{n-1}}\right) ,\text{ }U_{n}\triangleq\{\alpha_{i},\ |\alpha
_{i}|<\pi,i=1,...,4\},\text{ }\alpha_{i}\in\{\theta_{n-1}\pm|\cos^{-1}\Psi
_{n}|\mp2k\pi,\ k=0,1\}\ ,								\label{3.3b}
\end{equation}

\n {\it and the expectation integration is in the $(\bf a,{\it\F})$ region where, $\phi_n$ 
can only take the four values listed in the set $U_n$, in terms of the remaining 
integration variables.}

\svs\textit{Example 3} \ \ As an example of the implementation of the EGED formula (\ref%
{3.3}), consider the case $n=3$, and suppose $\phi _{i},\
i=1,2,3$ are independently uniformly distributed in $[-\pi,\pi]$, and the $A_{i}$%
s are constants. Then, application of formula (\ref{3.3}), a change of variable of integration 
and simplification, shows that if we let%
\begin{equation}
f_{\mathbf{A},\mathbf{\Phi}}(\mathbf{a},\mathit{\Phi})=(8\pi^{3})^{-1},\ \ \ 
\bar{b}_{2}^{2}\triangleq|\mathbf{A}|^{2}+2A_{1}A_{2}\cos (\phi_{1}),\ \ 
\text{and }\ \ \bar{\psi}_{3}\triangleq\left( \frac{b_{3}^{2}-\bar{b}%
_{2}^{2}-A_{3}^{2}}{2A_{3}\bar{b}_{2}}\right)\ ,  						\label{3.7}
\end{equation}
then we obtain \cite{Maghsoodi-1}
\begin{equation}
f_{B_{3}}(b_{3})=\left\{ 
\begin{array}{cc}
\nhs\nhs\tfrac{2b_{3}}{\pi^{2}}\int_{0}^{\pi}1_{\{\left\vert \bar{\psi}%
_{3}\right\vert \le 1\}}\frac{d\phi_{1}}{\sqrt{4A_{3}^{2}\bar{b}%
_{2}^{2}-(b_{3}^{2}-\bar{b}_{2}^{2}-A_{3}^{2})^{2}}} & m_{3}\leq b_{3}\leq
M_{3} \\ 
0\text{ \ \ \ \ \ \ \ \ \ \ \ \ \ \ \ \ \ \ \ \ \ \ \ \ \ \ \ \ \ \ \ \ \ \
\ \ \ \ \ \ \ \ \ \ } & \text{otherwise}.%
\end{array}
\right.   											\label{3.8}
\end{equation}
Graphs of numerical examples of this pdf are also illustrated in Figure 2 for various amplitude values.

Moreover, when $A_{i}=1,\ i=1,2,3$, the pdf (\ref{3.8}) further simplifies to \cite{Maghsoodi-1}
\begin{equation}
f_{B_{3}}(b_{3})=\left\{ 
\begin{array}{cc}
\nhs\nhs\frac{2b_{3}}{\pi^{2}}\int_{\tilde{m}_{2}}^{\tilde{M}_{2}}\frac{d\phi_{1}}{%
\sqrt{4\bar{b}_{2}^{2}-(b_{3}^{2}-\bar{b}_{2}^{2}-1)^{2}}} & 0\leq b_{3}\leq3
\\ 
0\text{ \ \ \ \ \ \ \ \ \ \ \ \ \ \ \ \ \ \ \ \ \ \ \ \ \ \ \ \ \ \ } & 
\text{otherwise}\ ,%
\end{array}
\right.  												\label{3.9}
\end{equation}
where in this case 

\n $\bar{b}_{2}^{2}\triangleq2+2\cos(\phi_{1})$, 
$\tilde {M}_{2}\triangleq\left\vert \cos^{-1}\left( (b_{3}-1)^{2}/2-1\right))\right\vert $, and $\tilde{m}_{2}\triangleq\left\vert \cos^{-1}(\min
(b_{3}+1,2)^{2}/2-1)\right\vert $. 

\n It follows from formula (\ref{3.9}) that in this case the pdf has a singularity at $b_3=1$ which can also be observed in the graph of Fig. 2.

\n Formula (\ref{3.9}) can also be written in terms of the EllipticF and EllipticK functions whose numerical values are
well tabulated and coded 
\begin{equation}
f_{B_{3}}(b_{3})=\frac{{4}Ib_{3}}{\pi^{2}}U\left\{ 
\begin{array}{cc}
\hskip-8mm\text{EllipticK}(P)-\text{EllipticF}(1/P,P) & 0\leq b_{3}\leq1 \\ 
\text{EllipticK}(P)\text{ \ \ \ \ \ \ \ \ \ \ \ \ \ \ \ \ \ \ \ \ \ \ \ \ \
\ \ \ \ } & 1<b_{3}\leq 3\ ,%
\end{array}
\right. \   												\label{3.10}
\end{equation}
where 
\begin{equation}
I\triangleq\sqrt{-1}\text{, }\ \ \ \ U\triangleq\frac{1}{\left\vert
1-b_{3}\right\vert \sqrt{(3+b_{3})(1-b_{3})}}\text{, }\ \ \ \ V\triangleq
(1+b_{3})^{-{3/2}}(3-b_{3})^{-\frac{1}{2}}\text{,}\ \ \ \ P\triangleq\frac {U%
}{V}\ ,  										\label{3.11}
\end{equation}
and%
\begin{equation}
\text{EllipticF}(Q,P)=\int_{0}^{Q} [(1-P^{2}x^{2})(1-x^{2})]^{-\frac{1}{2}%
}dx,\ \ \ \ \text{EllipticK}(P)=\text{EllipticF}(1,P)\ ,
\end{equation}

\n and formulae (\ref{3.9})-(\ref{3.11}) all give the required new
amplitude pdf. The pdf given in (\ref{3.9}) is illustrated in Fig. 2, which 
matches the pdf obtained from simulation, also illustrated in Fig. 2.

The pdf for the $n=2$ case of example 3, with different amplitudes $A_1$ and $A_2$, and independent uniform phases $\phi_1$ and $\phi_2$, can also be directly obtained from 
(\ref{3.7}) and (\ref{3.8}) above, by simply setting $A_1=0$, and replacing $b_3$ with $b_2$, $A_3$ with $A_2$, and $A_2$ with $A_1$, and noting that the integrand in (\ref{3.8}) becomes 
independent of $\phi_1$, hence giving 
\begin{equation}
f_{B_{2}}(b_{2})=\left\{ 
\begin{array}{cc}\hskip -2mm \frac{{2b_{2}}}
{{\pi \sqrt{4A_{1}^{2}A_{2}^{2}-(b_{2}^{2}-A_{1}^{2}-A_{2}^{2})^{2}}}} \hskip 20mm m_2 \le b_{2} \le M_2, \\ 
\hskip -4mm 0\text{\hskip 5.5cm otherwise.}%
\end{array} %
\right.  \label{3.11b}
\end{equation}
\n Examples of this pdf are also plotted in Figure 1, where the match with simulation is also observed.

\vskip 4mm\textit{Example 4} \ \ As another example of application of the EGED formula (\ref%
{3.3}), consider the $n=4$ case of example 1. Then, application of formula (\ref{3.3}), and again 
a change of variable of integration and simplification, shows that if we let    

\sn $f_{ {\bf A},{\bf \F} } ({\bf a},{\it \F})=(16\pi^4)^{-1},\ \ \bar b_3^2\den \ve {\bf A}\ve^2+2\bl[A_1A_2\cos(\f_1)+A_2A_3\cos(\f2)+A_1A_3\cos(\f_2-\f_1)\br]$,

\sn and \ \ $\bar\psi_4 \den {b_4^2-\bar b_3^2-A_4^2\ov 2A_4 \bar b_3}$,\ \ then we obtain \cite{Maghsoodi-1}
\begin{equation}
f_{B_{4}}(b_{4})=\left\{ 
\begin{array}{cc}
\tfrac{b_{4}}{2\pi^{3}}\int_{-\pi}^{\pi}\int_{-\pi}^{\pi}1_{\{\left\vert 
\bar{\psi}_{4}\right\vert \le 1\}}\frac{d\phi_{1}d\phi_{2}}{\sqrt{4A_{4}^{2}%
\bar{b}_{3}^{2}-(b_{4}^{2}-\bar{b}_{3}^{2}-A_{4}^{2})^{2}}} & m_{4}\leq
b_{4}\leq M_{4} \\ 
0\text{ \ \ \ \ \ \ \ \ \ \ \ \ \ \ \ \ \ \ \ \ \ \ \ \ \ \ \ \ \ \ \ \ \ \
\ \ \ \ \ \ \ \ \ \ \ } & \text{otherwise}\ ,
\end{array}
\right.   												\label{3.13}
\end{equation}
The pdf (\ref{3.13}) is also another hitherto undiscovered pdf, applicable in practical situations such as multiple 
access processes having one reference user and three interfering users, (see e.g. Fig. 4),  multipath channels 
with four taps, communication systems with cochannel interference, and so on. It is illustrated in Fig. 3, for $A_i=1,\ i=1,\cdots,4$. The graph 
matches that of  the pdf obtained from simulation, also illustrated in Fig. 3.  In similar fashions we can apply the EGED formula to derive the exact amplitude pdfs for
all other values of $n$, in terms of multiple integrals of order $n-2$ of functions of the given parameters.

\vskip 4mm\textit{Example 5}\ \ (Random Amplitudes) If the amplitudes are random with the joint density $f_{\bf A}(\bf a)$, and are independent of the phases, then it 
follows from formula (\ref{3.3}) that the envelope pdf, ${\tilde f}_{\tilde B}(\tilde b)$,  of this case, is the integral of the envelope pdf, $f_B  (b)$ of the corresponding constant amplitude case, 
w.r.t. the joint density of the amplitudes, i.e. 
\begin{equation}
{\tilde f}_{\tilde B}(\tilde b)=\int_{-\infty}^\infty\nhs\cdots \int_{-\infty}^\infty da_1\cdots  da_n\ f_{\bf A}({\bf a}) \ f_{B}(b)                     \label{3.14}
\end{equation}
For example the extension of the pdf obtained in (\ref{3.8}) of example 3 above, to the random amplitude case is 
\begin{equation}
{\tilde f}_{{\tilde B}_{3}}({\tilde b}_{3})=\tfrac{2b_{3}}{\pi^{2}} \int_{-\infty}^\infty \int_{-\infty}^\infty\int_{-\infty}^\infty\nhs da_1da_2da_3 f_{\bf A}({\bf a})  
\int_{0}^{\pi}  \frac{  1_{\{\left\vert { \tilde \psi}_{3}\right\vert \le 1\}}   d\phi_{1}}{\sqrt {4a_{3}^{2} \tilde {b}_{2}^{2}-(b_{3}^{2}-\tilde {b}_{2}^{2}-a_{3}^{2})^{2}}}\ \ \ m_{3}\leq b_{3}\leq M_{3}   
																				\label{3.15}
\end{equation}
\n where $\tilde b_2$ and $\tilde \psi_3$, respectively denote $\bar b_2$ and $\bar \psi_3$ of (\ref{3.7}), with $A_i$ replaced with $a_i$ for $i=1,2,3$. 

\n For example if the amplitudes ${\bf A}=(A_1,A_2,A_3)$ are jointly Gaussian, with mean vector $\bs {\mu}$ and covariance matrix $\bf C$, and the phases are distributed as in example 3, and are independent of the amplitudes, then the envelope density in this case is be given by (\ref{3.15}), with $f_{\bf A}(\bf a)$ substituted by 
\begin{equation}
f_{\bf A}({\bf a}) ={1 \ov {2\pi \sqrt {\ve {\bf C}\ve 2\pi} } } \exp \bl[ -\hf ({\bf a} -{\bs{\mu}})^T {\bf C}^{-1} ({\bf a} -{\bs{\mu}}) \br]				\label{3.16}
\end{equation}

\n For example in example 1 of section II, if the amplitudes are the same normally distributed random variable, with zero mean and variance $\sig^2$, then the pdf of the envelope is given by
\begin{align}
f_{B}(b)
&={2\ov \sig\pi\sqrt {2\pi} } \int_{-\infty}^\infty da  {\bf 1}_{\{\ve a\ve <b/2\}} { \exp -{1\ov 2\sig^2}a^2 \ov {\sqrt {4a^2-b^2} } } \cr
&={4\ov \sig\pi\sqrt {2\pi} } \int_{b/2}^\infty da {\exp -{1\ov 2\sig^2}a^2 \ov {\sqrt {4a^2-b^2} }} 	\cr
&={\sqrt 2\ov \sig\pi\sqrt {2\pi} } \int_0^2 dx {\exp -{b^2\ov 4\sig^2 x} \ov {x\sqrt {2-x} } }	\cr
&=-{1\ov \sig\pi\sqrt {2\pi} } e^{-{1\ov 4}c^2} K_0 \bl({1\ov 4}c^2\br)								\label{3.17}
\end{align}

\n where the third line in (\ref{3.17}) follows from the change of the variable of integration $a^2={b^2\ov 2x}$, and in the last line, $c\den {b\ov 2\sig}$, and $K_0$ denotes the modified Bessel function of the second kind of order zero.

\mvs\textit{Example 6} (Mixture of discrete and continuous distributions) The EGED formula 
(\ref {3.3}) can also be applied to the cases where the amplitudes and phases have an arbitrary mixture of 
discrete and continuous densities. To illustrate this power of the EGED formula, consider a two dimensional 
example where the amplitudes have an arbitrary joint continuous pdf $f(a_1,a_2)$, and are independent of the phases, and assume 
that the phases are mutually independent, taking only each of the discrete values $0$ or $\pi$, with probability $\hf$. 
Thus in this case the EGED formula becomes
\begin{equation}
f_{B}(b)=2bE\left\{ \left\vert 2A_2A_1\sqrt{1-\Psi_{2}}\right\vert^{-1}
1_{\{\left\vert \Psi_{2}\right\vert \le 1,\ \Phi_{2}\in U_{2}\}}\right\} 
\text{ \ \ }m_{2}\le b\le M_{2}\ ,  						\label{3.50}
\end{equation}
and $\phi_1$ replaces $\th_{n-1}$ in the definition (\ref {3.3b}) of the set $U_2$. Using the Dirac delta functions 
we can represent the discrete phase densities and write 
\begin{equation}
f_{B}(b)=2b\int_{-\infty}^{\infty}\int_{-\infty}^{\infty} { da_1da_2f(a_1,a_2)\ov \left\vert 2a_2a_1\sqrt{1-\psi_{2}}\right\vert}
{1\ov 4}\prod_{i=1}^2\int_{-\pi}^\pi d\phi_i\bl[ \del(\phi_i)+\del(\phi_i-\pi)\br]{\bf 1}_{\{\left\vert \psi_{2}\right\vert \le 1,\ \phi_{2}\in U_{2}\}} \label{3.51}
\end{equation}
Let $C_2\den \ve \cos^{-1}\psi_2\ve\in [0,\pi] $, and implement the delta functions $[ \del(\phi_1)+\del(\phi_1-\pi)\br]$ into the set $U_2$, followed by
implementing the delta functions $[ \del(\phi_2)+\del(\phi_2-\pi)\br]$ into the result. The product term in the latter part of (\ref{3.51}), which we denote 
by $\P$, will then become
\begin{align}
\P
&\den \prod_{i=1}^2 \int_{-\pi}^\pi d\phi_i   \bl[ \del(\phi_i)+\del(\phi_i-\pi)\br] {\bf 1}_{\{\ve \psi_{2} \ve \le 1,\ \phi_{2}\in U_{2}\}}\cr
&= \int_{-\pi}^\pi d\phi_2 \bl[ \del(\phi_2)+\del(\phi_2-\pi)\br] {\bf 1}_{\{\ve \psi_{2} \ve \le 1\}} \Bl[   {\bf 1}_{ \{\phi_2\in  \{\pm C_2,\ \pm(C_2-2\pi)\} \} } + 
{\bf 1}_{ \{ \phi_2\in \{\pi\pm C_2,\ \pi\pm(C_2-2\pi)\} \} }\Br ]\cr
&={\bf 1}_{\{\left\vert \psi_{2}\right\vert \le 1\} \}}  \Bl [ {\bf 1}_{\{C_2=0\}}+ {\bf 1}_{\{C_2=\pi\}} + {\bf 1}_{\{C_2=\pi\}} + {\bf 1}_{\{C_2=0\}}\Br] \cr
&={\bf 1}_{\{\left\vert \psi_{2}\right\vert \le 1\} \}}  \Bl [ {\bf 1}_{\{\psi_2=1\}}+ {\bf 1}_{\{\psi_2=-1\}} + {\bf 1}_{\{\psi_2=-1\}} + {\bf 1}_{\{\psi_2=1\}}\Br]        \label {3.52} 
\end {align} 
Substituting the last expression in (\ref{3.52}) back into (\ref{3.51}) whilst noting the denominator singularity at $\psi_2=\pm 1$ we have  
\begin{equation}
f_{B}(b)=2b\int_{-\infty}^{\infty}\int_{-\infty}^{\infty} da_1da_2{ f(a_1,a_2)\ov \left\vert 2a_2a_1\right\vert}
{1\ov 4}{\bf 1}_{\{\left\vert \psi_{2}\right\vert \le 1\} \}}  \Bl [ 2\del(\psi_2-1) + 2\del(\psi_2+1)\Br] 			\label {3.53}
\end{equation} 
Now using the formula  
\begin{equation}
\del (f(x))=\sum_{i=1}^n {\del(x-x_i) \ov \ve f'(x_i) \ve} \notag
\end{equation}
where the $x_i$ are the real roots of $f(x)$, the delta functions of $\psi_2$ can be written in terms of those of $a_2$, using the roots of $\psi_2=\pm 1$, which are $\{x_1=b-a_1,\ a_1\le b\}$, 
$\{x_2=a_1-b,\ a_1\ge b\}$, and $x_3=a_1+b$ respectively. We thus obtain the required envelope pdf from (\ref{3.53})
\begin{align}
f_{B}(b)
&=\hf b\int_{-\infty}^{\infty}\int_{-\infty}^{\infty} da_1  da_2 { f(a_1,a_2)\ov \left\vert a_2a_1\right\vert}
\sum_{i=1}^3{1\ov b}\Bl [\ve a_1 x_i\ve \del(a_2-x_i)\Br ]  \cr 			
&=\hf \Bl [\int_{-\infty}^b f(a_1,b-a_1)+ \int_b^{\infty} f(a_1,a_1-b) + \int_{-\infty}^{\infty} f(a_1,a_1+b)\Br] da_1			\label{3.54} 			
\end{align}
A simple special case of this example is when the amplitudes are independently negative exponentially distributed, with parameter $\lam$, thus allowing higher probabilities 
for lower amplitudes, we obtain an interesting envelope pdf 
$$f_{B}(b)=\hf \lam e^{-\lam b} ( \lam b + 1)$$

\vskip 6mm

\section{Application to BER performance}

As an application of the derived formulae, we consider the bit error rate
(BER) evaluation of BPSK systems over narrowband multipath fading channels.
In such channels, the delay spread $T_{m}$ of a channel is small relative to
the inverse signal bandwidth $B^{-1}$ of the transmitted signal, i.e. $%
T_{m}\ll B^{-1}$. This implies that the delay $\tau _{i}$ associated with
the $i$th multipath component $\tau _{i}\leq T_{m}$ $\forall $ $i$, so the
baseband signal $s(t)\approx s(t-\tau _{i})$ $\forall $ $i$. The received
signal representation with narrowband multipath fading and Gaussian noise in
the presence of $n$ multipath components is given by \cite{Goldsmith}\cite{Rappaport-book},
\begin{equation}
r(t)=\sum\limits_{i=1}^{n}\alpha _{i}\cos (2\pi f_{c}t+\phi _{i})+z(t),\text{%
\ \ \ \ \ \ \ \ }0\leq t<T_{s}  \label{E-r-Fading}
\end{equation}%
where $\alpha _{i}$ is the amplitude and $\phi _{i}$ is the phase of the $i$th
multipath component, respectively, the carrier frequency is denoted as $%
f_{c}$ and the symbol duration as $T_{s}$. In practice, the amplitudes $%
\alpha _{i}s$ of the individual multipath components do not fluctuate widely
over a local area because the channel characteristics change slowly with
respect to $T_{s}$. However, the phases $\phi _{i}s$ vary greatly even for
very small values of time delays because the distances traversed by the
propagating waves are orders-of-magnitude larger than the wavelength of the
carrier frequency. Therefore, the phases are usually modelled as independent
random variables uniformly distributed over $[-\pi ,\pi ]$ \cite{Proakis-1}-\cite{Rappaport-book}. In such channels, each individual term $\alpha
_{i}\cos (2\pi f_{c}t+\phi _{i})$ in (\ref{E-r-Fading}) is referred to as
a specular component.

Using the SAT formula, we can express (\ref{E-r-Fading}) as $r(t)=B_{n}\cos (\omega
t+\theta _{n})+z(t)$ where the amplitude $B_{n}$ is a random variable with a
pdf that depends on the value of $n$. A considerable simplification for the
channel model is achieved by using the common assumption that $n$ is large
and all amplitudes and phases are mutually independent, thus the Central
Limit Theorem can be invoked to approximate the received signal as a
Gaussian random process, hence the pdf of $B_{n}$ becomes the well-known
Rayleigh distribution. However, applying the Gaussian approximation (GA)
to the modelling of fading channels that have small number of specular
components, will not be accurate enough to describe the effects of the fading
process. Thus, the formulae derived in this work can be of great benefit to
solve the exact model of the narrowband fading channels, and evaluate the BER
of such channels significantly more accurately.

The average bit error probability $P_{B}$ for BPSK in narrowband multipath
fading channels can be obtained by integrating the corresponding error
probability in AWGN over the fading distribution \cite{Proakis-1}\cite{Goldsmith},
\begin{equation}
P_{B}=\int_{0}^{\infty }Q\left( \sqrt{2b_{n}^{2}\frac{E_{b}}{N_{0}}}\right)
\,f_{B_{n}}(b_{n})\text{ }db_{n}  \label{E-BER-Fading}
\end{equation}%
where $Q\left( \sqrt{2E_{b}/N_{0}}\right) $ is the probability of bit error
in AWGN channels. For large $n$, the pdf of $b_{n}$ is Rayleigh distributed
and (\ref{E-BER-Fading}) is reduced to a simple closed-form formula \cite%
{Proakis-1}. The computation of $P_{B}$ is usually performed versus the
average signal-to-noise ratio per bit $\bar{E}_{b}/N_{0}=E\left(
b_{n}^{2}\right) E_{b}/N_{0}$, where $E(b_{n}^{2})$ denotes the expected
value of $b_{n}^{2}$. For small $n$ values, the pdf of $b_{n}$ is given by (%
\ref{3.3}) which can be substituted in (\ref{E-BER-Fading}) to compute $P_{B}
$ for any $n$ value. The analytical and simulation results for $P_{B}$ as a
function of\ $\bar{E}_{b}/N_{0}$ are presented in Fig. $4$ for $n=1,2,3,4,$
the $n=2$ case with dependent phases is also included. All multipath
components are assumed to have equal average power. As demonstrated by Fig. $%
4$, the GA has a large discrepancy which is around $3$ dB for $n=3$, and it
is around $1.5$ dB for $n=4$. In the case of $n=2,$ which is known as the
Two-Ray model \cite{Goldsmith}, a large difference in $P_{B}$ is observed
between the dependent and the independent cases. Such behavior can be
understood with the aid of Fig. $1$ which shows that $P(b_{2}<1)$ with
independent phases is much larger than that with dependent phases, i.e., the
probability that the interference is destructive is much larger when the
phases are independent.

\mvs\section{Conclusion}

In this paper we considered the open problem of derivation of exact 
distributions of the envelopes of general stochastic sinusoidal sums, with random
amplitudes and phase angles, and its application in an important communication problem. 
We have seen that, in the most general case, it is possible to derive exact general formulae for the distribution 
of the resultant envelope in terms of just the given joint distribution of the amplitudes 
and the phase angles of the signals present in the sum. We derived two such general formulae, 
EDDHAPT and EGED, depending on the particular applications at hand. Examples 
showed that implementation of these formulae also lead to new explicit distributions 
which we applied to compute the exact BER performance of BPSK systems in narrowband fading
channels. The extension of these envelope pdf formulae 
to allow random number of signals is immediate under independence. Under no independence, 
the extension still follows naturally, but less immediately.

The presented formulae were applied to compute the exact BER performance of
BPSK systems in narrowband multipath fading channels with small number of
resolvable multipath components. All the simulation results were consistent with the 
exact theoretical findings, which also showed significant gain in accuracy over the currently 
popular Gaussian approximation method, particularly for small number of components.

In all cases, the formulae presented in this paper render themselves to
accurate and efficient numerical implementations, since at worst they merely
involve numerical computation of multiple integrals of known functions, the various 
algorithms for which are widely available and coded. In some cases the pdfs can be 
written in terms of known integrals such as Elliptic and Bessel
functions. Further numerical implementations and applications may also form
part of future work.

\section{Appendix}

\svs In this appendix we present the proofs of Theorem 1 and Theorem 2. 
These were first reported in \cite{Maghsoodi-1}, where details of the proofs and 
other related results may be found.

\mvs\subsection{Proof of Theorem {\rm 1 ( The EDDHAPT formula )}}

Formula (\ref{2.2}) can be written as%
\begin{equation}
B_{n}^{2}=\sum_{i=1}^{n}A_{i}^{2}+2A_{n}\sum_{k=1}^{n-1}A_{k}\cos(\phi
_{n}-\phi_{k})+2\nhs\sum_{(n-1)\geq j>k\geq1}\nhs A_{j}A_{k}\cos\left( \phi
_{j}-\phi_{k}\right)\ .  								\label{E_2.9}
\end{equation}
Expanding the $\cos$ terms and writing in terms of the half-angle tangents
we find that 

\begin{equation}
B_{n}^{2}=\sum_{i=1}^{n}A_{i}^{2}+2A_{n}\sum_{k=1}^{n-1} {{(1+T_nT_k)^2-(T_n-T_k)^2}\ov (1+T_n^2)(1+T_k^2) }A_{k}
+ 2\nhs\sum_{(n-1)\geq j>k\geq1}\nhs {{(1+T_jT_k)^2-(T_j-T_k)^2}\ov {(1+T_j^2)(1+T_k^2))}} A_{j}A_{k}	\label{E_2.9a}
\end{equation}

Hence collecting the $T_n$ terms in (\ref{E_2.9a}), we can write $B_{n}^{2}$ as a quadratic expression in $T_{n}$
\begin{equation}
Q\triangleq (1+T_n^2)(B_{n}^{2}-b_{n}^{2})=aT_{n}^{2}+4A_{n}\bar{S}%
_{n}T_{n}+(I_{n}+2A_{n}\bar{C}_{n})\ .  						\label{E_2.10}
\end{equation}
The probability on the LHS of (\ref{2.6}) is the multiple integral of the
pdf over the regions where the quadratic (\ref{E_2.10}) is non-positive. Thus distinguishing between the cases 
where $a$ or $\Delta_Q$ are positive or negative, we have that $Q$ is non-positive only in the union of three disjoint regions 
within the space $\{\mathbf{a},{\mathbf{t}}\}$, namely the regions 
$\{\Delta_Q>0, a>0, T_{n}\in\lbrack\bar{a}_{n},\bar{b}_{n}]\}$, $\{\Delta_Q>0, a<0,$ $T_{n}\notin\lbrack\bar{b}_{n},\bar{a}_{n}]\}$, and $\{\Delta_Q<0, a<0\}$ 
, where $\bar{a}_{n}$ and $\bar{b}_{n}$ are the roots of $Q$, and $\bar b_n$ is the larger root in the first region, and vice-versa in the second. Hence
\begin{align}
P(B_n \leq b_n) =&\int_{-\infty}^\infty\nhs d{\bf a} \Biggl\{ \int_{-\infty}^{\infty}\nhs  d\bar {\bf t} {\bf 1}_{\{\Delta_Q>0, a>0\}} \int_{\bar a_n}^{\bar b_n}\nhs dt_n f_{{\bf A},{\bf T}} ( {\bf a, t})
+  \int_{-\infty}^{\infty}\nhs d\bar {\bf t} {\bf 1}_{\{\Delta_Q>0, a<0\}}   
\Bl\{ \int_{-\infty}^{\bar b_n}\nhs +\int_{\bar a_n}^{\infty} \Br\} dt_n f_{{\bf A},{\bf T}} ( {\bf a, t})\cr
&\shs + \int_{-\infty}^{\infty}\nhs d\bar {\bf t} {\bf 1}_{\{\Delta_Q<0, a<0 \}}   f_{{\bf A},\bar {\bf T}} ( {\bf a}, \bar{\bf t}) \Biggr\}\label{E_2.10b}
\end{align}
The second term in (\ref{E_2.10b}), which we denote by $I_2$, can be written as 
\begin{align}
I_2
&=\int_{-\infty}^{\infty}\nhs  d{\bf a} \Biggl\{ \int_{-\infty}^{\infty}\nhs d\bar {\bf t} {\bf 1}_{\{\Delta_Q>0, a<0 \}}  \int_{-\infty}^{\infty} dt_n f_{{\bf A},{\bf T}} ({\bf a, t})
- \int_{-\infty}^{\infty}\nhs  d\bar {\bf t} {\bf 1}_{\{\Delta_Q>0, a<0 \}}  \int_{\bar b_n}^{\bar a_n}\nhs dt_n f_{{\bf A},{\bf T}} ( {\bf a, t})\Biggr\}  \cr
&=\int_{-\infty}^{\infty}\nhs  d{\bf a} \Biggl\{ \int_{-\infty}^{\infty}\nhs d\bar {\bf t} {\bf 1}_{\{\Delta_Q>0, a<0 \}}   f_{{\bf A},\bar {\bf T}} ( {\bf a}, \bar{\bf t})
+  \int_{-\infty}^{\infty}\nhs   d\bar {\bf t} {\bf 1}_{\{\Delta_Q>0, a<0 \}} \int_{\bar a_n}^{\bar b_n}\nhs dt_n f_{{\bf A},{\bf T}} ( {\bf a, t}) \Biggr\}    \label{E_2.10c}
\end{align}
Substituting the RHS of (\ref{E_2.10c}), into (\ref{E_2.10b}) for $I_2$, and combining the integrals, we obtain (\ref{2.6}) and the proof of the Theorem is complete  \bull

\svs {\it Remark}\ \ Note that in the regions where $a=0$\ and/or\ $\Del_Q=0$, the probability measures are zero, hence these cases need not be included in the regions of integration.

\bvs\subsection{Proof of Theorem {\rm 2 (The EGED formula )} }

\svs The proof is by a method which we call {\it The} ($n-1$)-{\it Conditioning Method}, where we calculate the cdf of $B_n^2$ by conditioning on the value of the random vector $(\bf A, \bar {\bf\F})$, where $\bar {\bf\F}$ denotes the vector $({\bf \F}_1,...,{\bf \F}_{n-1})$.  We obtain 
\begin{equation}
P\bl(B^2_n\le b_n^2\br)
=\int_{-\infty}^{\infty}\nhs d{\bf a}\int_{-\pi}^{\pi} d\ibF\ P\Bl (\cos(\F_n-\th_{n-1})\le \psi_n  / {\bf a}, \ibF\Br) f_{_{_{ {\bf A},\bar {\bf \F}}}}({\bf a}, \ibF)\ ,   \label{E_2.11}
\end{equation}

\n where we have used the EST formula (\ref{3.1}), ${\it \bar\F}$ denotes the vector $(\f_1,...,\f_{n-1})$, and without loss of generality $a_{n}$ is assumed to be positive (see the
remark below), and 
\begin{equation}
\psi_n\den \Bl( {b_n^2-b_{n-1}^2-a_n^2\ov 2a_n b_{n-1}}\Br)\ .						\label{E_2.12}
\end{equation}
 
\n Taking inverse $\cos$ within the probability integrand of (\ref{E_2.11}), while noting that $(\F_n-\th_{n-1})\in (-2\pi, 2\pi)$, we obtain  
\begin{equation} P\bl(B^2_n\le b_n^2\br) = \int_{-\infty}^{\infty} \nhs d{\bf a} \int_{-\pi}^{\pi}\nhs d\ibF P\Bl (\ve\cos^{-1} \bar\psi_n\ve +\ve \th_{n-1}\ve \le \ve \F_n\ve \le 2\pi-\ve\cos^{-1} \bar\psi_n\ve +\ve \th_{n-1}\ve / {\bf a}, \ibF\Br) f_{_{_{{\bf A},\bar {\bf \F}}}}({\bf a}, \ibF)					\label{E_2.13}
\end{equation}
\vskip -3mm\n where $\F_n$ and $\th_{n-1}$ carry the same sign, and the function 
\begin{equation}
\bar\psi_n =\left\{ 
\begin{array}{cc}
\psi_n & |\psi_n|\leq 1
\\ 
\hbox{sign} (\psi_n) & \text{otherwise}\ ,
\end{array}
\right. \notag
\end{equation}
\n ensures that the $\cos^{-1}$ function is well-defined.  Note that, as required by (\ref{E_2.11}), when $\psi_{n}\ge 1$ or $\le -1$, the RHS of (\ref{E_2.13}) takes the values 1 and $0$ respectively. Now writing (\ref{E_2.13}) in terms of the conditional cdf $F_{_{\F_n/{\bf A},\bar{\bf \F}}}(\phi_n)$ of $\F_n$, given that $({\bf A},\bar {\bf \F})=({\bf a},  \ibF)$, which we simply denote by $\tilde F_n$, and letting $C_k(\bar\psi_n)\den \ve\cos^{-1}\bar  \psi_n\ve-2k\pi$, we obtain  
\begin{equation}
P\bl(B^2_n\le b_n^2\br) 
=\int_{-\infty}^{\infty}\nhs d{\bf a}\int_{-\pi}^{\pi}\nhs d\ibF \Bl [\sum_{k=0}^1 \bl\{\tilde F_n(-C_k(\bar\psi_n)+\th_{n-1}) -\tilde F_n(C_k(\bar\psi_n)+\th_{n-1})\br\} \Br ] 
f_{_{_{{\bf A}, \bar {\bf \F}}}} ({\bf a}, {\ibF} )\ .													\label{E_2.14}
\end{equation}

\n While noting that, $f_{B_n}(b_n)=2b_n f_{B_n^2}(b_n^2)$, we differentiate (\ref{E_2.14}) with respect to $b_n^2$ to get the pdf. The derivative of $\bar\psi_n$ w.r.t. $b_n^2$ yields the function  $\tilde{\bf 1}\den {\bf 1}_{\bl\{\ve\psi_n\ve\le 1\br\}}$ into the integrand, which is followed by $\tilde f_n$,  the conditional pdf of $\F_n$
\begin{equation}
f_{B_n}(b_n)  =2 b_n \intf\nhs d{\bf a}\int_{-\pi}^{\pi}\nhs d\ibF \ \tilde {\bf 1} \Bl [ \sum_{k=0}^1 \bl\{\tilde f_n(-C_k(\psi_n)+\th_{n-1}) 
+ \tilde f_n(C_k(\psi_n)+\th_{n-1})\br \} \Br]   V(\psi_n) f_{_{_{{\bf A}, \bar {\bf \F}}}} ({\bf a}, {\ibF} )			\label{E_2.15}
\end{equation}
\n where $V(\psi_n)\den (2a_n b_{n-1} \sqrt  { 1- {\psi}_n^2 }\br)^{-1} =- {\partial \ve\cos^{-1}(\psi_n)\ve\over\partial b^2_n}$. Note that $\bar\psi_n$ no longer needs to replace $\psi_n$ within the integrands in (\ref{E_2.15}), since now the function $\tilde {\bf 1}$ forces the integrand to zero when $\psi_n$ is outside the range $[-1,1]$, which is as required by the fact that the probability sum in (\ref{E_2.14}) takes the constant values of 1 or 0, for all $\psi_n\notin [-1,1]$ (see also (\ref{E_2.11})), hence its derivative within the integrand of (\ref{E_2.15}) should be zero in this region. The product of $\tilde f_n$ and the marginal pdf $f_{_{_{{\bf A}, \bar {\bf \F}}}} ({\bf a}, {\ibF} )$ in the integrand of (\ref{E_2.15}), gives the joint pdf $f_{ {\bf A},{\bf \F} } ({\bf a},{\it \F})$, hence we have 
\begin{align} f_{B_n}(b_n)  
&=2 b_n \intf\nhs d{\bf a}\int_{-\pi}^{\pi}\nhs d\ibF \ \tilde {\bf 1} \Bl [ \sum_{k=0}^1 \bl\{ f_{ {\bf A},{\bf \F} } ({\bf a},{\ibF},-C_k(\psi_n)+\th_{n-1}) 
+ f_{ {\bf A},{\bf \F} } ({\bf a},{\ibF},C_k(\psi_n)+\th_{n-1}) \br \} \Br]   V(\psi_n) \cr
&=2 b_n \intf\nhs d{\bf a}\int_{-\pi}^{\pi}\nhs d{\it \F}\ \tilde {\bf 1} \Bl [ \sum_{k=0}^1 \Bl\{ {\bf 1}_{\bl\{\f_n=-C_k(\psi_n)+\th_{n-1}\br\}}
+ {\bf 1}_{\bl\{\f_n=C_k(\psi_n)+\th_{n-1}\br\}}\Br\}\Br ]  V(\psi_n)f_{ {\bf A},{\bf \F} } ({\bf a},{\it \F})\ .				\label{E_2.16}
\end{align}
\n Finally writing (\ref{E_2.16}) in terms of the expectation with respect to the given joint law $f_{_{ {\bf A},{\bf \F} }} ({\bf a},{\it \F})$ we obtain (\ref{3.3}) and the proof of Theorem 2 is complete.\hfill \bull

\svs\textit{Remark} \ \ \ Note that identical steps prove Theorem 2 for the region of integration where $a_{n}$ is negative, as 1 minus the probability in the integrand
of (\ref{E_2.14}) is obtained, which after differentiation gives the same result with a negative sign, which changes to the positive sign, by replacing $a_{n}$ with $-|a_n|$ in front of the function $V(\psi_n)$. Hence, for all $a_{n}$, (\ref{E_2.16}) holds with, $|a_{n}|$ replacing $a_{n}$ in front of $V(\psi_n)$, hence formula  (\ref{3.3}) follows.

\vfill\eject

\begin{figure}
\centering
\includegraphics[width=4.5in]{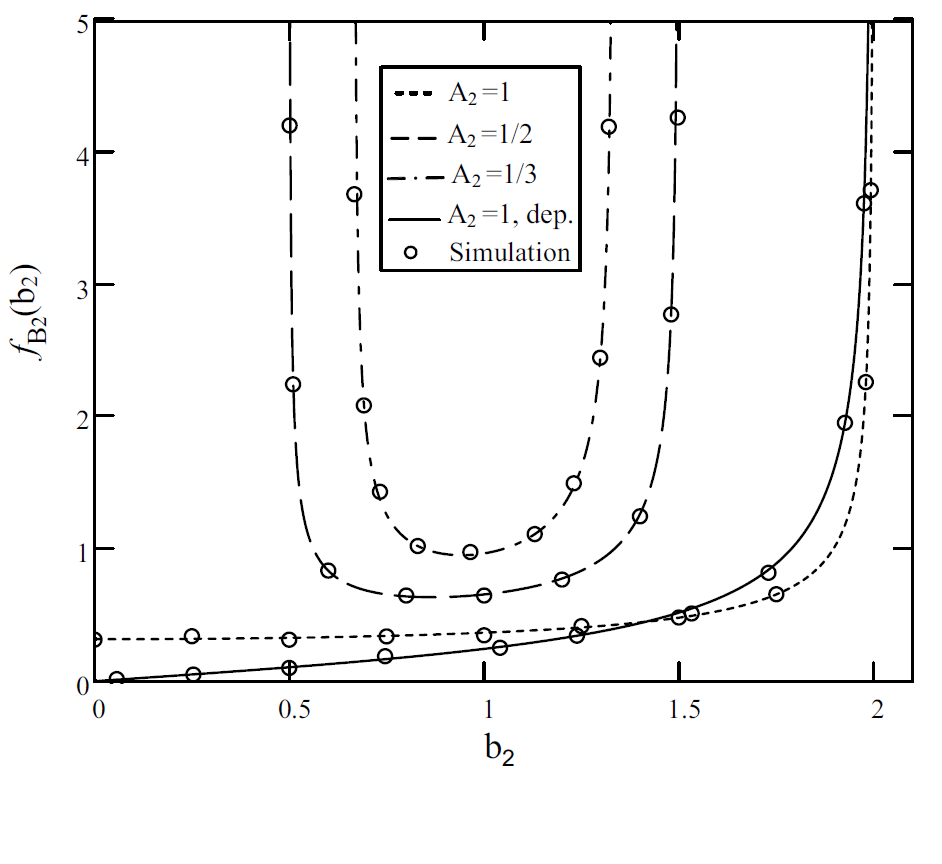}\setlength{\belowcaptionskip}{-13pt} 
\caption{Plot of the pdf's of examples 1 to 3, $n = 2$ with dependent and independent phases, $A_1=1$.}
 \label{fig-1}
\end{figure}

\begin{figure}
\centering
\includegraphics[width=4.5in]{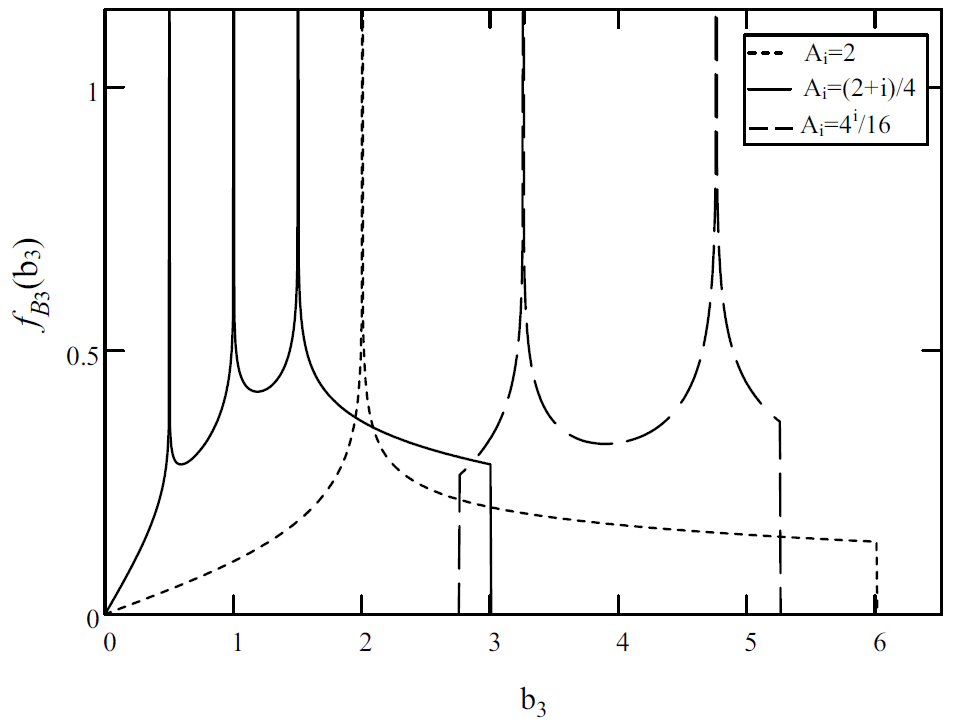}\setlength{\belowcaptionskip}{-13pt} 
\caption{Plot of the pdf's of example 3, $n=3$, $i=1, 2, 3$.}
 \label{fig-1}
\end{figure}

\begin{figure}
\centering
\includegraphics[width=4.5in]{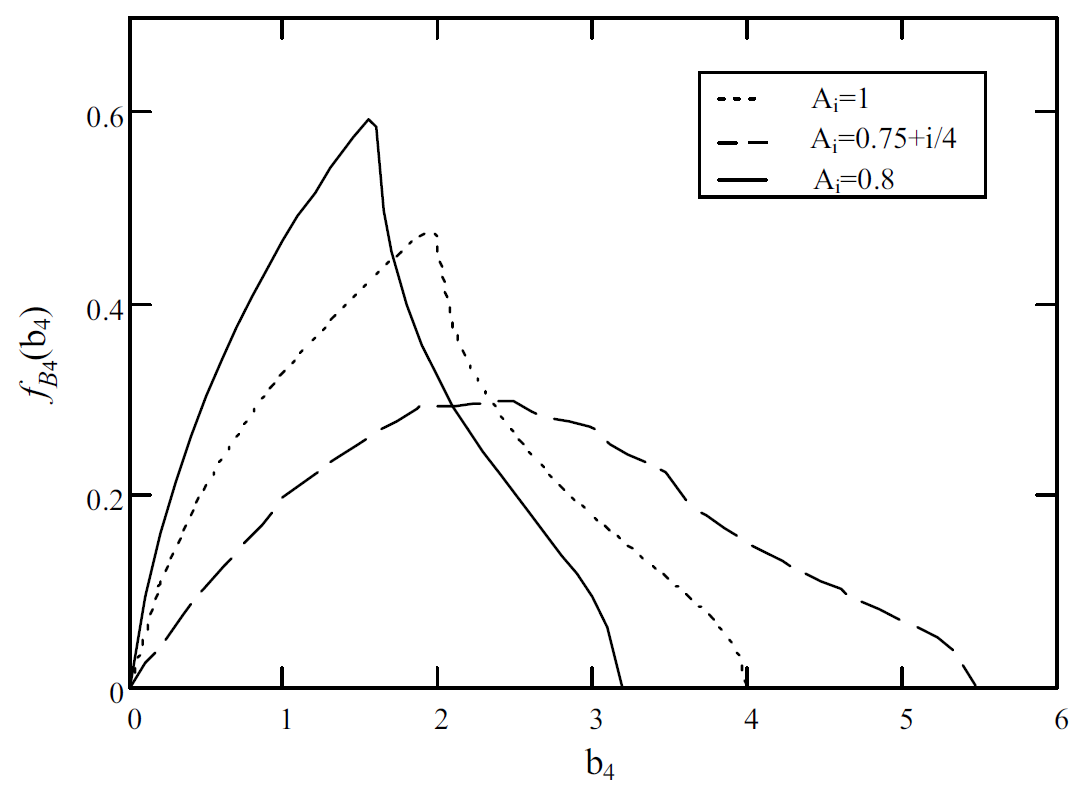}\setlength{\belowcaptionskip}{-13pt} 
\caption{Plot of the pdf's of example, $n=4$, and $i=1,2,3,4$.}
 \label{fig-1}
\end{figure}

\begin{figure}
\centering
\includegraphics[width=4.5in]{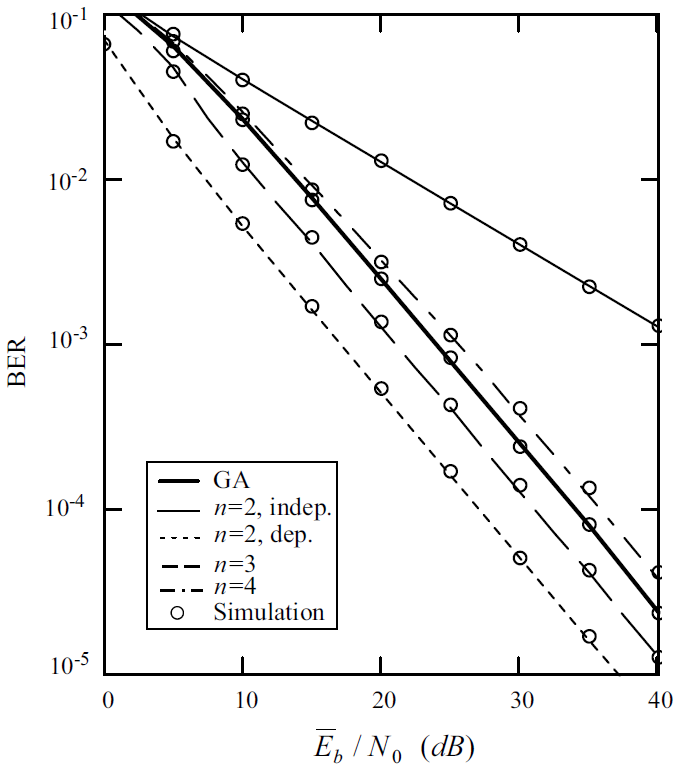}\setlength{\belowcaptionskip}{-13pt} 
\caption{Probability of error for different number of multipath components, via the exact formulae, Gaussian Approximation and simulation.}
 \label{fig-1}
\end{figure}

\begin{figure}
\centering
\includegraphics[width=6.5in]{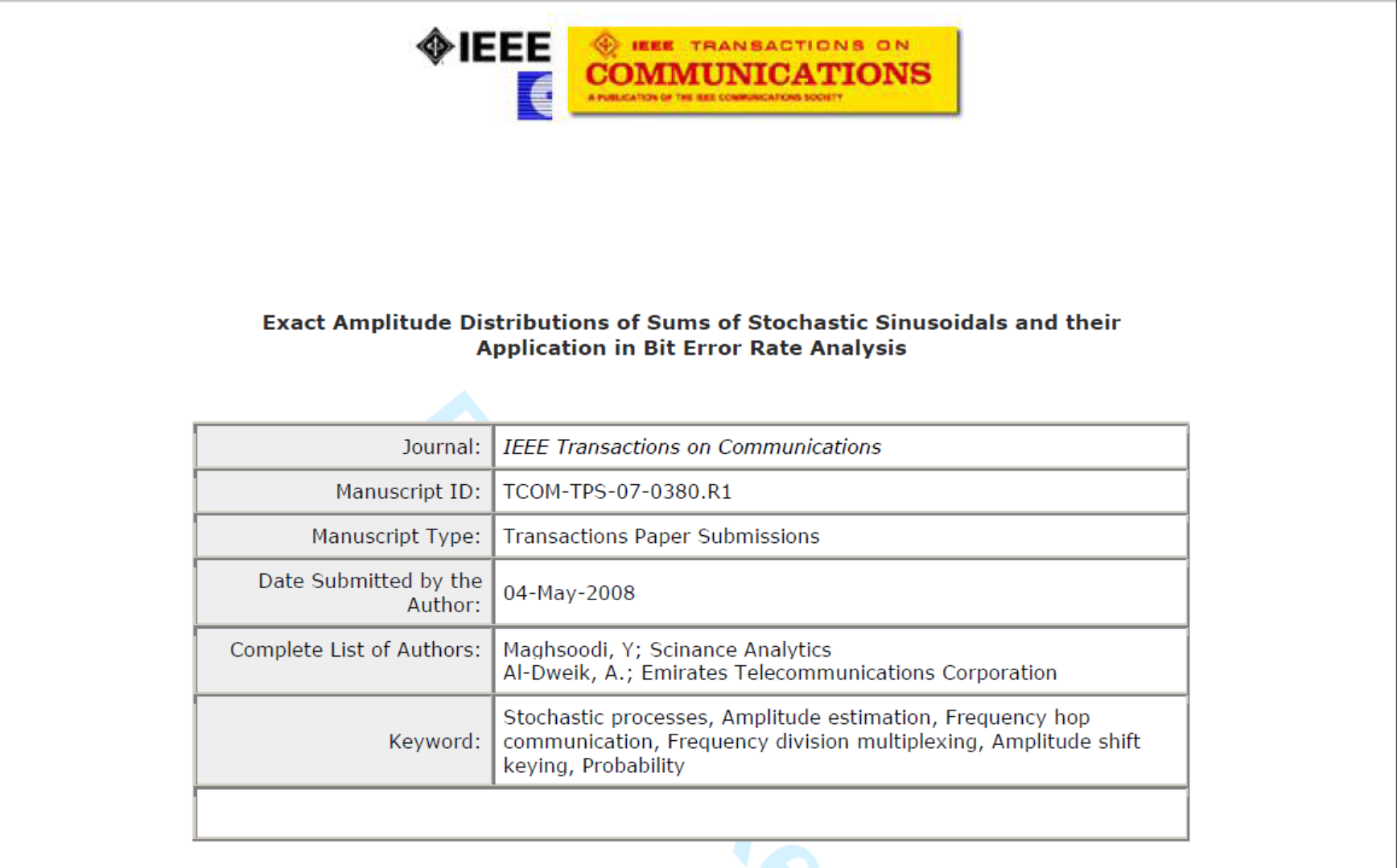}\setlength{\belowcaptionskip}{-13pt} 
\caption{Probability of error for different number of multipath components, via the exact formulae, Gaussian Approximation and simulation.}
 \label{fig-1}
\end{figure}


\begin{thebibliography}{99}
\bibitem{Proakis-1} J. G. Proakis, \textit{Digital Communications},
McGraw-Hill, 2001.

\bibitem{Goldsmith} A. J. Goldsmith, \textit{Wireless Communications, }New
York, Cambridge University Press, 2005.

\bibitem{Rappaport-book} T. S. Rappaport, \textit{Wireless Communications
Principles and Applications}, New Jersy, Prentice Hall, 2002.

\bibitem{Kah-1} K. Teh, A. Kot, and K. Li, \textquotedblleft Multitone
jamming rejection of FFH/BFSK spread-spectrum system over fading
channels,\textquotedblright\ \textit{IEEE Trans. Commun.}, vol. 46, , pp.
1050-1057, Aug. 1998.

\bibitem{Kwonhue-1} K. Choi an K. Cheun, \ \textquotedblleft Performance of
asynchronous slow frequency-hop multiple-access networks with MFSK
modulation,\textquotedblright\ \textit{IEEE Trans. Commun.}, vol 48, pp.
298-307, Feb. 2000.

\bibitem{Arafat-1} A. Al-Dweik and F. Xiong, \textquotedblleft
Frequency-hopped multiple access communications with noncoherent M-ary
OFDM-ASK, \textit{IEEE Trans. Commun.}, vol. 51, pp. 33-36, Jan. 2003.

\bibitem{Beaulieu-1} N. C. Beaulieu and J. Cheng, \textquotedblleft Precise
error-rate analysis of bandwidth-efficient BPSK in Nakagami fading
channels,\textquotedblright\ \textit{IEEE Trans. Commun.}, vol. 52, pp.
149-158, Jan. 2004.

\bibitem{Kim 1} S. H. Kim and S. W. Kim, \textquotedblleft Frequency-hopped
multiple-access communications with multicarrier on-off keying in Rayleigh
fading channels,\textquotedblright\ \textit{IEEE Trans. Commun.,} vol. 48,
pp. 1692-1701, Oct. 2000.

\bibitem{Rice-1} S. O. Rice, \textquotedblleft Probability distributions for
noise plus several sine waves--The problem of computation," \textit{IEEE
Trans. Commun.}, COM-22, pp. 851-853, Jun. 1974.

\bibitem{Kluyver} J. Kluyver, \textquotedblleft A local probability
problem,\textquotedblright\ Nederlande Akademie van Wetenschap, 8, pp.
341-350, 1906.

\bibitem{Simon-1} M. K. Simon, \textquotedblleft On the probability density
function of the squared envelope of a sum of random
vectors,\textquotedblright\ \textit{IEEE Trans. Commun.}, COM-33, pp.
993-996, Sept. 1985.

\bibitem{Helstrom-1} C. W. Helstrom, \textquotedblleft Distribution of the
sum of two sine waves and Gaussian noise,\textquotedblright\ \textit{IEEE
Trans. Inform. Theory}, vol. 38, pp. 186-191, Jan. 1992.

\bibitem{Helstrom-2} C. W. Helstrom, \textquotedblleft Distribution of the
envelope of a sum of random sine waves and Gaussian
noise,\textquotedblright\ \textit{IEEE Trans. Aerosp. Electron. Syst.}, vol.
35, pp. 594-601, Apr. 1999.

\bibitem{Beckmann-2} P. Beckmann and A. Spizzichino, \textit{The Scattering
of Electromagnetic Waves From Rough Surfaces}, 2nd ed. Boston, MA: Artech
House, 1987.

\bibitem{Zabin} S. M. Zabin and G. A. Wright, \textquotedblleft
Nonparametric density estimation and detection in impulsive interference
channels-Part I: Estimators,\textquotedblright\ \textit{IEEE Trans. Commun.}%
, vol. 42, pp. 1684-1697, 1994.

\bibitem{Abdi} A. Abdi, H. Hashemi, and S. Nader-Esfahani, \textquotedblleft
On the PDF of the sum of random vectors,\textquotedblright\ \textit{IEEE
Trans. Commun.,} vol. 48, pp. 7-12, Jan. 2000.

\bibitem{Maghsoodi-1} Y. Maghsoodi "Exact distributions of envelopes of sums of stochastic sinusoids with general random amplitudes and phases," Working paper, Scinance Analytics, Nov. 2004.

\bibitem{Arafat-1} Y. Maghsoodi and A. Al-Dweik, "Error-Rate Analysis of FHSS Networks Using Exact Envelope Characteristic Functions of Sums of Stochastic Signals," in IEEE Transactions on Vehicular Technology, vol. 57, no. 2, pp. 974-985, March 2008.

\bibitem{Arafat-2} A. Al-Dweik and B. Sharif, "Exact Performance Analysis of Synchronous FH-MFSK Wireless Networks," in IEEE Transactions on Vehicular Technology, vol. 58, no. 7, pp. 3771-3776, Sept. 2009.


\bibitem{Arafat-3} A. Al-Dweik, B. Sharif and C. Tsimenidis, "Accurate BER Analysis of OFDM Systems Over Static Frequency-Selective Multipath Fading Channels," in IEEE Transactions on Broadcasting, vol. 57, no. 4, pp. 895-901, Dec. 2011.



\end{thebibliography}
\end{document}